\documentclass{appolb}
\usepackage{amsmath}
\usepackage{amsfonts}
\usepackage{amssymb}
\usepackage{xcolor}
\usepackage{color}
\usepackage{float}
\usepackage[english]{babel}
\usepackage{graphics}
\usepackage{graphicx}
\begin{document}
\title{Nonclassical behavior of moving relativistic \\
unstable  particles
}
\author{K. Urbanowski\footnote{e--mail: K.Urbanowski@if.uz.zgora.pl}
\address{Institute of Physics,
University of Zielona G\'{o}ra,  \\
ul. Prof. Z. Szafrana 4a,
65-516 Zielona G\'{o}ra,
Poland}}

\maketitle
\begin{abstract}
We study the survival probability of moving relativistic
unstable particles with definite momentum $\vec{p} \neq 0$.
The amplitude of the survival probability
of these particles
is calculated using its integral representation.
We found decay curves of such particles for the quantum
mechanical models considered. These model studies show that late
time deviations of the survival probability of these particles
from the exponential form of the decay law, that is  the
transition times region between exponential
and non-expo\-nen\-tial form of the survival
probability, should  occur much earlier than it follows from the classical standard
approach resolving itself into
replacing time $t$ by $t/\gamma$
(where $\gamma$ is the relativistic Lorentz factor) in the formula
for the survival probability
and that the survival probabilities should
tend to zero as $t\rightarrow \infty$ much slower than one
would expect using classical time dilation relation.
Here we show also
that for some physically admissible models of unstable states the
computed decay curves of the moving particles have
fluctuating form at relatively short times including times of  order of the lifetime.
\end{abstract}


\section{Introduction}

Physicists studying the decay processes are often confronted with the problem of how
to predict the form of the decay law  of the  particle moving in respect  to
the rest reference frame of the observer knowing the decay law
of this particle decaying in its rest frame.
From the standard, text book  considerations one finds that if
the decay law of the unstable particle in rest
 has the exponential form ${\cal P}_{0}(t) = \exp\,[- \frac{{\it\Gamma}_{0}\,t}{\hbar}]$
then the decay law of the moving particle with momentum $p \neq0$ is
${\cal P}_{p}(t)
\,= \,\exp\,[-\,\frac{{\it\Gamma}_{0}\,t}{\hbar \,\gamma}]$, where $t$ denotes time,
${\it\Gamma}_{0}$ is the decay rate (time $t$ and ${\it\Gamma}_{0}$
are measured in the rest reference frame of the particle)
and $\gamma$ is the
relativistic Lorentz factor, $\gamma \equiv 1 / {\sqrt{1 - \beta^{2}}},
\;\;\beta = v/c$, $v$ is the velocity of the particle.
This equality is the classical physics relation.
It is almost common belief that this equality is valid also for any $t$
in the case of quantum decay processes
and does not depend on the model of the unstable particles considered.
For the proper interpretation of many accelerator experiments with high
energy unstable particles as well as of results of observations of astrophysical
processes in which a huge numbers of elementary particles
(including unstable one) are produced
we should be sure that this belief is supported by theoretical analysis of
quantum models of decay processes. The problem seems to be extremely important
because from some theoretical studies it follows that in the case of quantum
decay processes this relation is valid to a sufficient accuracy only for not
more than a few lifetimes
 $\tau_{0} = \hbar / {\it\Gamma}_{0}$ \cite{stefanovich,shirkov,exner,ku-2014}.
What is more it
appears that this relation may not apply in the case of the famous result of
the GSI experiment, where an oscillating decay rate of the ionized isotopes
$^{140}$Pr and $^{142}$Pm moving with relativistic velocity
($\gamma \simeq 1.43$) was observed \cite{litvinov1,kienle1}.
So we can see that the problem requires a deeper  analysis.
In this paper
the basis of such an analysis
 will be the formalism developed in \cite{stefanovich,shirkov}
where within the quantum field theory the formula for
the survival amplitude of moving particles was derived.
We  will follow the method used in \cite{ku-2014} and
we will analyze
numerically properties of the survival probability for
a model of the unstable particle based on the Breit--Wigner
mass distribution  considered therein and  as well as the other  different one.
Here we show that the relativistic treatment of the problem
within the Stefanovich--Shirokov theory \cite{stefanovich,shirkov} yields decay curves
tending to zero as $t\rightarrow \infty$ much slower than one would expect using
classical time dilation relation which confirms
and generalizes some conclusions drawn in \cite{ku-2014}.
We show also that for some physically admissible models of unstable states
decay curves of the moving particles
computed using the above mentioned approach have
analogous
fluctuating form
as the decay curve  measured in the GSI experiment
and that
in the model considered
these fluctuations begin from times much shorter than the lifetime.
Our results shows that conclusions relating to the quantum decay
processes of moving particles based on the use of the classical physics
time dilation relation  need not be universally valid.

One of the aims
of this paper is
to analyze numerically properties of the survival
probability
in a wide range of times $t$ from very short
$t \ll \tau_{0}$ trough $t \sim \tau_{0}$ until $t \gg \tau_{0}$
of moving unstable particles derived in \cite{stefanovich,shirkov}
and to present results of calculations of
decay curves of such particles
for the model considered in \cite{ku-2014}
but for the more realistic parameters  of this model and to confront them  with
results obtained for another more realistic model.
Another intention is to demonstrate that when considering the relativistic quantum unstable
system the only rational assumption seems to be the assumption that the momentum $\vec{p}$ of
such a system is constant.
The paper is organized as follows:  Sec. 2 contains preliminaries and
the main steps of the derivation of all relations
necessary for the numerical studies, which results are presented in
Sec. 3.  Consequences of the assumption that
the momentum $\vec{p}$ of the moving freely quantum unstable
system is constant are analyzed in Sec. 4.
Sec. 5 contains a discussion and conclusions.

\section{Decay law of moving relativistic particles}

Let us begin our considerations from the following assumptions:
Suppose that in a laboratory a large number ${\cal N}_{0}$ of
unstable particles was created at the instant of time $t_{0}$ and then
their decay process is observed there.
Suppose also that all these unstable  particles do not move or are
moving very slowly in relation to the rest frame of the reference
 of the observer ${\cal O}$
 and  that  the observer counts at
instants $t_{1} < t_{2} < \ldots t_{n} < \ldots$, (where $t_{1} > t_{0}$),
how many particles ${\cal N}(t)$ survived up to these instants of time. All
collected results of these observations can be approximated by a function of time
${\cal P}_{0}(t) \simeq {\cal N}(t) / {\cal N}_{0}$ forming a decay curve.
If ${\cal N}_{0}$ is large then ${\cal P}_{0}(t)$ can be considered as
the survival probability of the unstable particle. The standard procedure is
to confront results of the experiment with theoretical predictions. Within
the quantum theory, when one intends to analyze the survival probability
${\cal P}_{0}(t)$ of the unstable state or particle, say $\phi$, in the rest system,
one starts from the calculation of the probability amplitude $a_{0}(t)$.
This amplitude defines the survival probability
${\cal P}_{0}(t) = |a_{0}(t)|^{2}$ we are looking for.
There is $a_{0}(t) \equiv \langle \phi|\phi(t)\rangle$  and
$|\phi (t)\rangle = \exp\,[-\frac{i}{\hbar} t H]\,|\phi\rangle$,
where $H$ is the total, self--adjoint Hamiltonian of the system considered.
Studying the properties of the amplitude  $a_{0}(t)$  it is convenient to use
the integral representation of $a_{0}(t)$ as the Fourier transform of the energy or,
equivalently mass distribution function, $\omega(m)$,
(see, eg. \cite{Fock,khalfin-2,fonda,muga,muga-1,calderon}),
with $\omega (m) \geq 0$ and $\omega (m) = 0$ for $m < \mu_{0}$, ($\mu_{0}$
is the lower bound of the spectrum of $H$). It appears that the general form of
the decay law ${\cal P}_{0}(t)$ of the particle in its rest reference frame practically
does not depend on the form of the all physically
acceptable $\omega(m)$ (see, eg. \cite{nowakowski3,fonda,muga,muga-1,urbanowski-1-2009,urbanowski-2017,giraldi}):
There is, $a_{0}(t) = a_{exp}(t) + a_{lt}(t)$, starting from times
slightly longer than the extremely short times \cite{urbanowski-1-2009,urbanowski-2017,giraldi}. Here
 $a_{exp}(t) = N\,\exp\,[{-i\frac{t}{\hbar}\,(E_{0} - \frac{i}{2}\,{\it\Gamma}_{0})}]$,
($E_{0}=m_{0}\,c^{2}$ is the energy of the system in the unstable state
$|\phi\rangle$ measured at the canonical decay times when ${\cal P}_{0}(t)$
has the exponential form, $N$ is the normalization constant). The component $a_{lt}(t)$
exhibits inverse power--law behavior at the late time region. The late
time region denotes times $t > T$, where $T$ is the cross--over time and it
can be found by solving the following equation, $|a_{exp}(t)|^{\,2} = |a_{lt}(t)|^{\,2}$.
There is $|a_{exp}(t)| \gg |a_{lt}(t)|$ for $t < T$
and $|a_{exp}(t)| \ll |a_{lt}(t)|$ for $t > T$.

We came to the place where a flux  of  moving relativistic
unstable particles investigated by an observer in his
laboratory should be considered.
According to the fundamental principles of the classical physics and quantum
theory (including relativistic quantum field
theory) the energy and momentum of the
moving particle have to be conserved. There is no
an analogous conservation law for the velocity $\vec{v}$.
These conservation laws are one of the
basic and  model independent tools of the study
of reactions between the colliding or decaying particles.
So it seems to be reasonable to assume,
as it was done in \cite{stefanovich,shirkov,khalfin-1},
that momentum $\vec{p}$ of the moving unstable particles measured
in the rest frame of  the observer is constant
(see also a discussion  in \cite{giacosa1}).
The question is what is the picture seen by the observer in such a case
and what is the relation between this picture and the picture seen by this
observer in the case of non moving unstable particles? In other words we should compare
the decay law ${\cal P}_{0}(t)$ with the decay law
${\cal P}_{p}(t)$ of the moving relativistic unstable particle
with the definite, constant momentum $\vec{p}=\,$const..
It is important to remember that the decay law ${\cal P}_{p}(t)$ does not describe the quantum
decay process of the moving particle in its rest frame but describes
the decay process of this particle seen by the observer in his rest laboratory frame.
Such a picture one meets in numerous experiments in the field of high energy physics or
when detecting cosmic rays: Detectors of a finite volume
are stationary in the frame of reference of the observer ${\cal O}$ and stable or unstable
particles together with their decay products passing through the detector are recorded.
According to the broadly shared opinion reproduced in many textbooks one
expects that  it should be,
\begin{equation}
{\cal P}_{p}(t) = {\cal P}_{0}(t/\gamma), \label{a-p=a-0}
\end{equation}
in the considered case.
This relation is a simple extension  of the  standard time dilation formula
to quantum decay processes.
The question is how does the time dilation formula
being the classical physics formula
work in the case of quantum decay processes?
From the results reported in \cite{stefanovich,shirkov,ku-2014} and
obtained there for the model defined by Breit--Wigner mass (energy) distribution
function $\omega(m) = \omega_{BW}(m)$ it follows that the relation (\ref{a-p=a-0}).
works in this model only within a limited range of times: For no more than a few lifetimes
What is more, it has been shown in \cite{ku-2014} that for times longer than few
lifetimes the difference between the correctly obtained survival
probability ${\cal P}_{p}(t)$ and ${\cal P}_{0}(t/\gamma)$
is significant.

Now let us follow \cite{stefanovich,shirkov} and
calculate  survival probabilities ${\cal P}_{0}(t)$ and ${\cal P}_{p}(t)$.
Hamiltonian $H$ and the momentum operator ${\bf P}$ have
common eigenvectors $|m;p\rangle$.
Momentum $\vec{p}$ is the eigenvalue of the momentum
operator ${\bf P}$. There is in $\hbar = c =1$ units:
\begin{equation}
{\bf P}|m;p\rangle = \vec{p}|m;p\rangle, \label{m;P}
\end{equation}
and
\begin{equation}
H|m;p\rangle = E'(m,p)\,|m;p\rangle. \label{H-m;p}
\end{equation}
In the coordinate system of the unstable quantum state
at rest, when $\vec{p} =0$, we have $|m;0\rangle = |m;p=0\rangle$,
\begin{equation}
H|m;0\rangle = m\,|m;0\rangle, \label{H-m;0}
\;\;\;m\in \sigma_{c}(H),
\end{equation}
where $m \equiv E'(m,0)$ and  $\sigma_{c}(H)$ is the continuous
part of the spectrum of the Hamiltonian  $H$.
Operators $H$ and ${\bf P}$ act in the state space ${\cal H}$.
Eigenvectors $|m;p\rangle$ are normalized as follows
\begin{equation}
\langle p;m|m';p\rangle = \delta(m - m'). \label{d-m-m'}
\end{equation}

Now we can model the moving unstable
particle $\phi$ with constant momentum,
$\vec{p}$, as the  following wave--packet $|\phi_{p}\rangle$,
\begin{eqnarray}
 |\phi_{p}\rangle
= \int_{\mu_{0}}^{\infty}\varsigma(m)\, |m;p\rangle\,dm,\label{phi-p}
\end{eqnarray}
where
expansion coefficients $\varsigma(m)$ are functions of the mass parameter $m$, that is of the rest mass $m$,
which is Lorentz invariant and therefore the scalar functions $\varsigma(m)$ of $m$ are
also Lorentz invariant.
(Here  $\mu_{0}$ is the lower bound of the spectrum $\sigma_{c}(H)$ of $H$).
We  require the state $|\phi_{p}\rangle$  to be normalized:
So it has to be $\int_{\mu_{0}}^{\infty}|\varsigma(m)|^{2}\,dm = 1$.

By means of the relation  (\ref{phi-p}) we can define the state vector
$|\phi\rangle \stackrel{\rm def}{=}  |\phi_{0}\rangle
\equiv |\phi_{p=0}\rangle \in {\cal H}$  describing
an unstable state in rest as follows,
\begin{eqnarray}
|\phi_{0}\rangle = |\phi \rangle =
\int_{\mu_{0}}^{\infty}\,\varsigma(m) |m;0\rangle\, dm. \label{phi}
\end{eqnarray}
This expansion and (\ref{H-m;0}) allow one  to find the amplitude $a_{0}(t)$
and to write
\begin{equation}
a_{0}(t)  \equiv \int_{\mu_{0}}^{\infty} \omega(m)\;
e^{\textstyle{-\,i\,m\,t}}\,d{m},
\label{a-spec}
\end{equation}
where $\omega(m) \equiv  |\varsigma(m)|^{2} > 0$.

We need also the probability amplitude
$a_{p}(t) = \langle \phi_{p}|\phi_{p} (t)\rangle$,
which defines the survival probability ${\cal P}_{p}(t) = |a_{p}(t)|^{2}$.
There is $|\phi_{p} (t) \rangle \stackrel{\rm def}{=}
\exp\,[-itH]\,|\phi_{p}\rangle$ in $\hbar = c = 1$ units.
We have the vector $|\phi_{p}\rangle$ (see (\ref{phi-p}))
but we still need eigenvalues $E'(m,p)$ solving  Eq. (\ref{H-m;p}).
Vectors $|\phi\rangle, |\phi_{p}\rangle$ are
elements of the same state space ${\cal H}$
connected with the coordinate rest system of the observer ${\cal O}$:
We are looking for the decay law of the moving
particle measured by the observer ${\cal O}$.
If to assume for simplicity that ${\bf P} = (P_{1},0,0)$ and that
$\vec{v} = (v_{1},0,0) \equiv (v,0,0)$ then there is $\vec{p}=(p,0,0)$ for
the eigenvalues $\vec{p}$ of the momentum operator ${\bf P}$.
Let $\Lambda_{p,m}$ be the Lorentz transformation from the reference
frame ${\cal O}$, where the momentum of the unstable particle considered is zero,
$\vec{p}=0$, into the frame ${\cal O}'$ where  the momentum of this
particle is $\vec{p} \equiv (p,0,0) \neq 0\;$ and $p \geq 0$,
or, equivalently, where its velocity equals $\vec{v} = \vec{v}_{p,m} \equiv
\frac{\vec{p}}{m \,\gamma_{m}}$, (where $m$ is the rest mass and
$\gamma_{m} \equiv \sqrt{p^{2} + (m)^{2}}/m$).
In this case the corresponding
4--vectors are: $\wp =(E/c,0,0,0) \equiv (m,0,0,0) \in {\cal O}$ within
the considered system of units, and $\wp'=(E'/c,p,0,0)
\equiv (E',p,0,0) = \Lambda_{p,m}\;\wp \,\in \,{\cal O}'$.
There is  $\wp'\cdot \wp' \equiv (\Lambda_{p,m}\; \wp)\cdot (\Lambda_{p,m}\; \wp)
= \wp\cdot \wp$ in Minkowski space, which is an effect  of the  Lorentz
invariance. (Here the dot "$\cdot$" denotes the scalar product in Minkowski space).
Hence, in our case: $\wp'\cdot \wp' \equiv (E')^{2} - p^{2} = m^{2}$
because $\wp \cdot \wp \equiv m^{2}$ and
thus $(E')^{2} \equiv (E'(m,p))^{2} = {p^{2} + m^{2}}$.

Another way to find $E'(m,p)$ is to use
the unitary representation, $U(\Lambda_{p,m})$, of the transformation
$\Lambda_{p,m}$, which acts  in the Hilbert space ${\cal H}$ of states
$|\phi\rangle \equiv |\phi;0\rangle, |\phi_{p}\rangle \,\in\,{\cal H}$:
One can show that the vector $U(\Lambda_{p,m}) |m;0\rangle$ is the common
eigenvector for operators $H$ and $\bf P$, that is that there is
\[
|m;p\rangle \equiv U(\Lambda_{p,m})|m;0\rangle
\]
 (see, eg. \cite{gibson}). Indeed,
taking into account that operators $H$ and  $\bf{P}$ form a 4--vector
$P_{\nu}=(P_{0},{\bf P}) \equiv (P_{0},P_{1},0,0)$, and $P_{0}  \equiv H$, we have
\[
U^{-1}(\Lambda_{p,m}) P_{\nu}U(\Lambda_{p,m})
= \Lambda_{p,m;\;\nu \lambda}\,P_{\lambda},
\]
where $\nu, \lambda = 0,1,2,3$ (see, e.g., \cite{gibson}, Chap. 4).
From this general transformation rule it follows that
\begin{eqnarray}
U^{-1}(\Lambda_{p,m}) P_{0} U(\Lambda_{p,m}) &=& \gamma_{m}\,(P_{0}\, +\,
v_{m}\,P_{1}) \nonumber \\
& \equiv & \gamma_{m}(H\, +\, v_{m}\,P_{1}). \label{U-H-U}
\end{eqnarray}
Based on this relation, one can show that
that vectors $U(\Lambda_{p,m})|m;0\rangle$  are
eigenvectors for the Hamiltonian $H$. There is
\begin{eqnarray}
H \,U(\Lambda_{p,m})|m;0\rangle
&=& U(\Lambda_{p,m})\,U^{-1}(\Lambda_{p,m})\,H\,U(\Lambda_{p,m})|m;0\rangle \nonumber \\
&=& \gamma_{m}\,U(\Lambda_{p,m})\, (H\, + \,v_{m}\, P_{1})\,|m;0\rangle. \label{h;m-1}
\end{eqnarray}
The Lorentz factor $\gamma_{m}$ corresponds to the rest mass $m$ being the
eigenvalue of the vector $|m;0\rangle$.
There are
$\gamma_{m} \neq \gamma_{m'}$ and $v_{m} \neq v_{m'}$ for  $m \neq m'$.
From (\ref{m;P}) it follows that $P_{1}\;|m;0\rangle = 0$ for $p = 0$,
which means that using (\ref{H-m;0}) the relation (\ref{h;m-1})
can be rewritten as follows
\begin{eqnarray}
H\,U(\Lambda_{p,m})|m;0\rangle = m \gamma_{m}\,U(\Lambda_{p,m})|m;0\rangle.
 \label{h;m-2}
\end{eqnarray}
Taking into account the form of the $\gamma_{m}$
forced by the condition $p =$ const one concludes
that in fact the eigenvalue found, $m\gamma_{m}$,
equals $m \gamma_{m} \equiv \sqrt{p^{2} + m^{2}}$.
This is exactly the same result as
that at the conclusion following from the
Lorentz invariance mentioned earlier: $E'(m,p) = \sqrt{p^{2} + m^{2}}$, which
shows   that the above considerations are self--consistent.

Similarly one can show that vectors $U(\Lambda_{p,m})|m;0\rangle$ are the
eigenvectors of the momentum operator ${\bf P}$
for the eigenvalue $m\gamma_{m}\, v_{m}\equiv p$,
that is that $U(\Lambda_{p,m})|m;0\rangle \equiv |m;p\rangle$ which was to show.

Thus finally we come to desired result:
\begin{equation}
H|m;p\rangle = \sqrt{p^{2} + m^{2}}\,|m;p\rangle \label{H;m-p}
\end{equation}
which replaces Eq. (\ref{H-m;p}).

Now using (\ref{H;m-p}) and the equation (\ref{phi-p})
we obtain the final, required  relation for the amplitude
$a_{p}(t)$,
\begin{eqnarray}
a_{p}(t)
&=& \int_{\mu_{0}}^{\infty} \omega(m)\;
e^{\textstyle{-\,i\sqrt{p^{2}+m^{2}}\;\,t}}\,d{m}. \label{a-p}
\end{eqnarray}
The above derivation of the expression for $a_{p}(t)$
is similar to that of \cite{ku-2014}. It is based on \cite{gibson}
and
it is reproduced here for the convenience of readers.
This is a shortened and
slightly changed, simplified version of the considerations presented in
\cite{stefanovich} and mainly in \cite{shirkov} and more explanations and more
details can be found therein and in \cite{stefanovich1,shirkov1},
where this formula was  derived using the quantum field theory  theory approach.

\section{Results of numerical studies}

\begin{figure}[h!]
\begin{center}
\includegraphics[width=88mm]{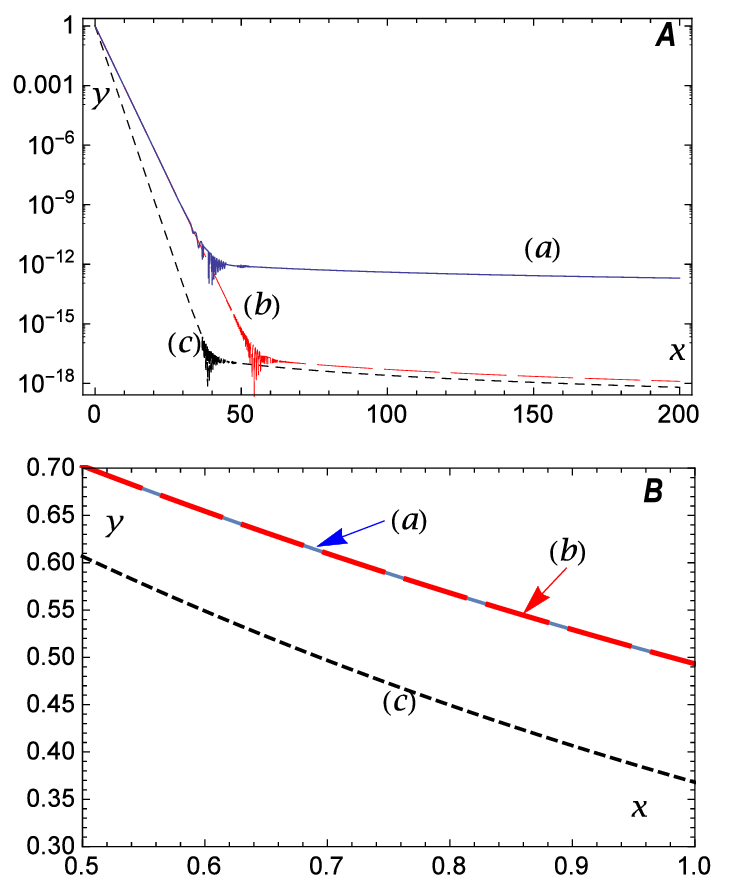}\\
\caption{Decay curves obtained for $\omega_{BW}(m)$ given by Eq. (\ref{omega-BW}).
Axes: $x =t / \tau_{0} $ --- time $t$ is measured in lifetimes
$\tau_{0}$,   $y$ --- survival probabilities (panel $A$: the logarithmic
scales, $(a)$ the decay curve ${\cal P}_{p}(t)$, $(b)$ the decay curve
${\cal P}_{0}(t/\gamma)$, $(c)$ the decay curve  ${\cal P}_{0}(t)$; panel
$B$:  $(a)$ -- ${\cal P}_{p}(t)$, $(b)$ --  ${\cal P}_{0}(t/\gamma)$, $(c)$ -- ${\cal P}_{0}(t)$ ). }
  \label{f1}
\end{center}
\end{figure}

\begin{figure}[H]
\begin{center}
\includegraphics[width=88mm]{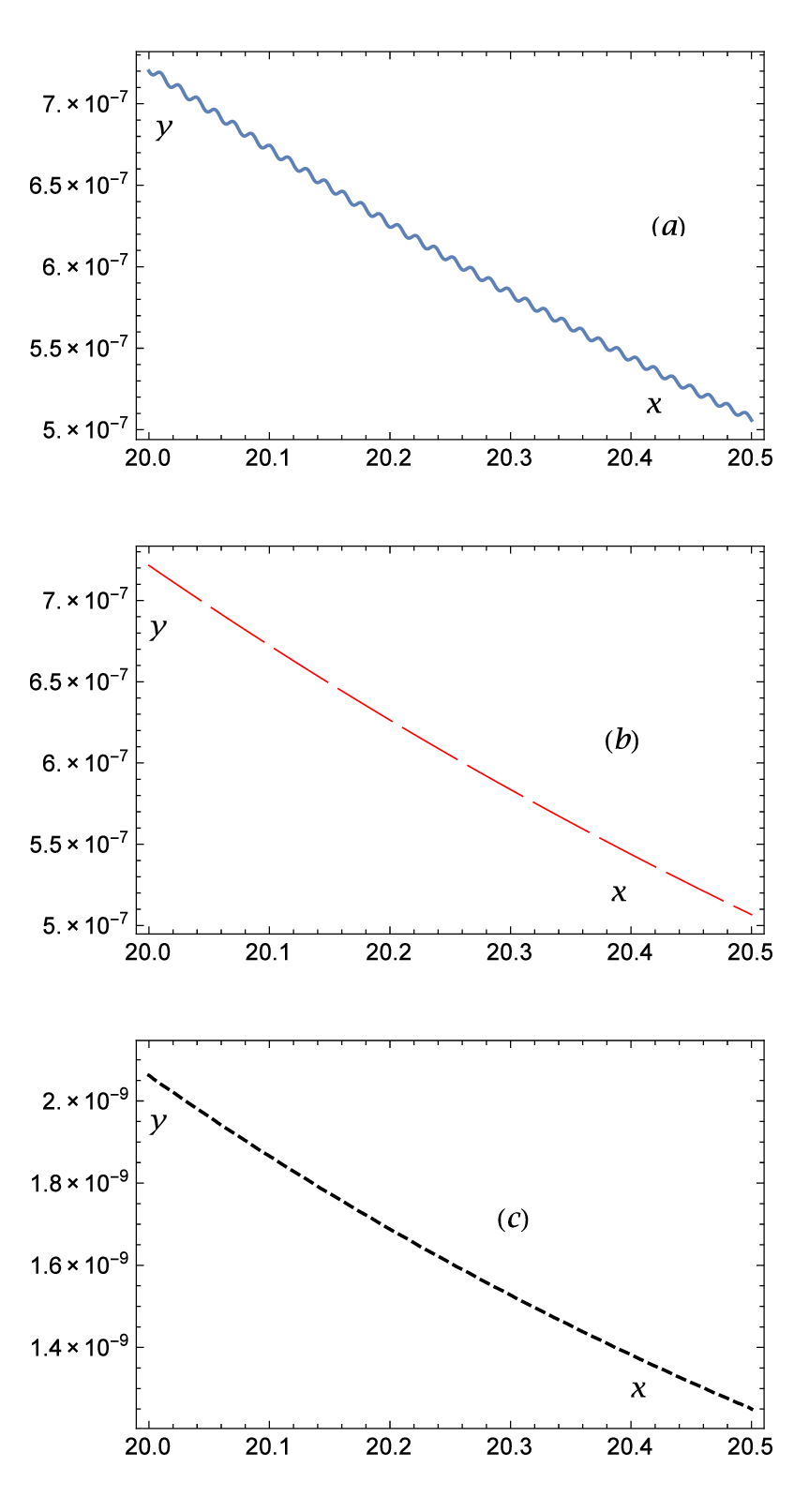}\\
\caption{Decay curves obtained for $\omega_{BW}(m)$ given by Eq. (\ref{omega-BW}).
Axes: $x =t / \tau_{0} $, and  $y$ -- survival probabilities:
 $(a)$ the decay curve ${\cal P}_{p}(t)$, $(b)$ the decay curve
${\cal P}_{0}(t/\gamma)$, $(c)$ the decay curve ${\cal P}_{0}(t)$. }
  \label{f2}
\end{center}
\end{figure}

 According to the literature a reasonable simplified
 representation of the density of the mass
distribution is to choose the Breit--Wigner form
$\omega_{BW}(m)$ for $\omega (m)$,  which under
rather general condition approximates sufficiently well many real
systems \cite{stefanovich,fonda,nowakowski3},
\begin{equation}
\omega_{BW}(m) \stackrel{\rm def}{=} \frac{N}{2\pi} {\it\Theta} (m - \mu_{0})
\frac{{\it\Gamma}_{0}}{(m - m_{0})^{2} +
(\frac{{\it\Gamma}_{0}}{2})^{2}}, \label{omega-BW}
\end{equation}
where $N$ is a normalization constant and ${\it\Theta} (m)$ is the
unit step function, $m_{0}$ is the rest mass of the particle and ${\it\Gamma}_{0}$
is the decay rate of the particle in the rest. Inserting
$\omega (m) \equiv \omega_{BW}(m)$ into (\ref{a-spec}) and into
(\ref{a-p}) one can find decay curves (survival probabilities)
${\cal P}_{0}(t)$ and ${\cal P}_{p}(t)$.
Results of numerical calculations are presented in Figs (\ref{f1}), (\ref{f2})
where calculations were performed for $\mu_{0} = 0$, $E_{0}/{\it\Gamma}_{0}
\equiv m_{0}/{\it\Gamma}_{0} = 1000$
and $cp/{\it\Gamma}_{0} \equiv p/{\it\Gamma}_{0} = 1000$. Values of these
parameters correspond to $\gamma = \sqrt{2}$, which is very close to $\gamma$
from the experiment performed by the GSI team \cite{litvinov1,kienle1} and
this is why such values of them were chosen in our considerations.
Similar calculations were performed in \cite{ku-2014} but for different and less
realistic values of the ratio $ m_{0}/{\it\Gamma}_{0} $:
For $ m_{0}/{\it\Gamma}_{0} = 10, \;25$ and $100$ and different $p/{\it\Gamma}_{0}$.
According to the literature
for laboratory systems a typical
value of the ratio $m_{0}/{{\it\Gamma}_{0}}$ is
$m_{0}/{{\it\Gamma}_{0}} \,\geq \, O (10^{3} - 10^{6})$ (see eg. \cite{krauss}) therefore the
choice $m_{0}/{\it\Gamma}_{0} = 1000$ seems to
be reasonable and more realistic than those used in \cite{ku-2014}.

Results presented in
Figs (\ref{f1}) and (\ref{f2})
show that in the case of $\omega (m)$
having the Breit--Wigner form the survival probabilities ${\cal P}_{0}(t/\gamma)$ and
 ${\cal P}_{p}(t)$  overlap for not too long times when
${\cal P}_{p}(t)$ has the canonical, that is the exponential form.
This observation confirms conclusions drawn in \cite{stefanovich,shirkov,ku-2014}.
On the other hand results presented in the panel $A$ of Fig (\ref{f1}) and Fig
(\ref{f2}) show that in the case of moving relativistic unstable
particles the transition times region, when
the canonical form of the survival probability ${\cal P}_{p}(t)$ transforms
into inverse power like form of $t$, begins much earlier than in the case of this
particle observed in its rest coordinate system and described by ${\cal P}_{0}(t)$.
This observation agrees with  results obtained in \cite{ku-2014}.

To be sure that the above conclusions are valid not only in
the approximate case $\omega_{BW}(m)$
of the density  of the mass distribution $\omega(m)$
we should consider a more general form of $\omega (m)$.
The most general condition for $\omega (m)$ following from (\ref{a-spec}) is that $\omega (m) \in \; L_{1}(-\infty, \infty)$.
So, if to assume that $\omega (m) \in \; L_{1}(-\infty, \infty)$ and additionally that $\omega (m) = 0$ for $m < \mu_{0}$,  $\omega (\mu_{0} =0)$ and $\omega (m)  \geq 0$ for $m> \mu_{0}$, that is that
\begin{equation}
\omega ({ m}) = {\it\Theta} (m - \mu_{0})\,( { m} - { \mu}_{0})^{\iota + l}\;
\varrho ({ m}),
\label{omega-eta}
\end{equation}
(where $0 \leq \iota < 1, l=0,1,2,\ldots;$), and
$\varrho (\mu_{0}) \stackrel{\rm def}{=}
\varrho_{0} > 0$, $\varrho (m) \geq 0$ for $m > \mu_{0}$ and $\varrho^{(k)}({ m}) = \frac{d}{dm}\,\varrho(m)$,
($k= 0,1,\ldots, n$),
exist and they are continuous
in $[{\mu}_{0}, \infty)$, and  limits
$\lim_{ m \rightarrow {\mu}_{0}+}\;
\varrho^{(k)}( m) \stackrel{\rm def}{=} \varrho_{0}^{(k)}$ exist, and
\[
\lim_{{ m} \rightarrow \infty}\,
( { m} - { \mu}_{0})^{\iota +l}\,\varrho^{(k)}({ m}) = 0,
\]
for all above mentioned {$k$}, then
one finds for $l=0$ that in the rest system (see \cite{urbanowski-1-2009,urbanowski-2017}),
\begin{eqnarray}
a_{0}(t) & \begin{array}{c}
          {} \\
          \sim \\
          \scriptstyle{t \rightarrow \infty}
        \end{array} &
        (-1)\,e^{\textstyle{-{i}{ \mu}_{0} t}}\;
        \Big[
        \Big(- \frac{i}{t}\Big)^{\iota + 1}\;
        \Gamma(\iota + 1)\;\varrho_{0}\; \label{a-eta} \\
        && +\;\iota\,\Big(- \frac{i}{t}\Big)^{\iota + 2}
        \;\Gamma(\iota + 2)\;\varrho_{0}^{(1)}\;+\;\ldots
        \Big]
         = a_{lt}(t),
            \nonumber
\end{eqnarray}
where $\Gamma(x)$ is Euler's gamma function. Hence,  one finds that, e.g. for $\iota = 1/2$, the leading term of $a_{lt}(t)$ has the following form
\begin{eqnarray}
a_{lt}(t) & \simeq &
        (-1)\,e^{\textstyle{-{i}{ \mu}_{0} t}}\; \frac{\sqrt{\pi}}{2}\;
        \Big[
        \Big(- \frac{i}{t}\Big)^{3/2}\;
        \varrho_{0}\;+\ldots \Big]. \label{a-eta-1/2}
\end{eqnarray}

\begin{figure}[h!]
\begin{center}
\includegraphics[width=88mm]{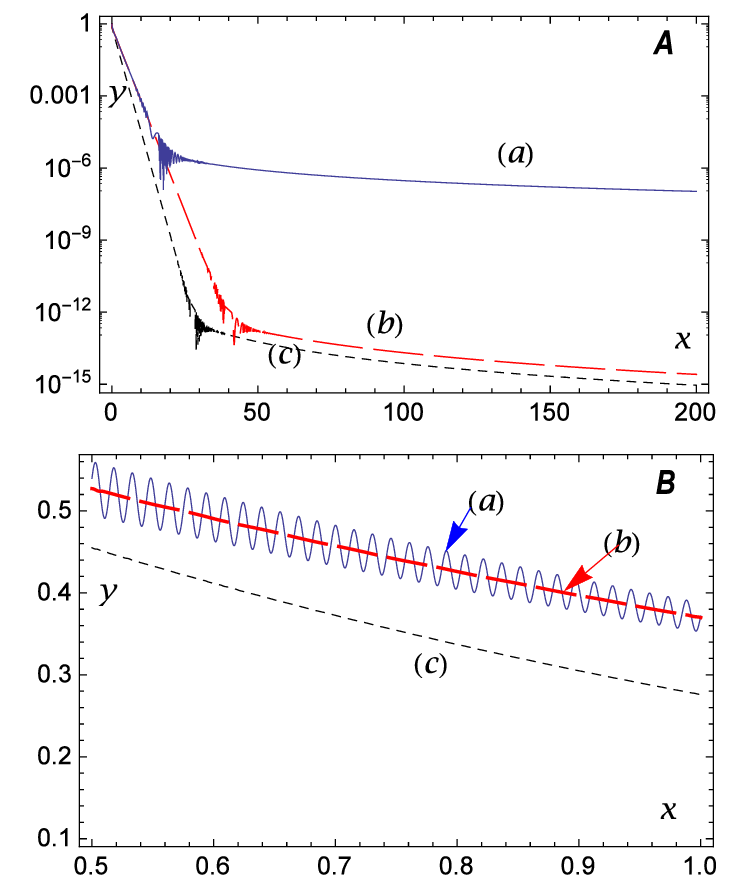}\\
\caption{Decay curves obtained for $\omega(m)$ given by Eq. (\ref{omega-exp}).
Axes: $x =t / \tau_{0} $, and $y$ --- survival probabilities
(panel $A$: the logarithmic scales, $(a)$ the decay curve
${\cal P}_{p}(t)$, $(b)$ the decay curve ${\cal P}_{0}(t/\gamma)$,
$(c)$ the decay curve ${\cal P}_{0}(t)$; panel $B$:
$(a)$ -- ${\cal P}_{p}(t)$, $(b)$ --  ${\cal P}_{0}(t/\gamma)$, $(c)$ -- ${\cal P}_{0}(t)$ ). The case $\eta {\it\Gamma}_{0}/m_{0} = 0.01$.}
  \label{f3}
\end{center}
\end{figure}

\begin{figure}[h!]
\begin{center}
\includegraphics[width=88mm]{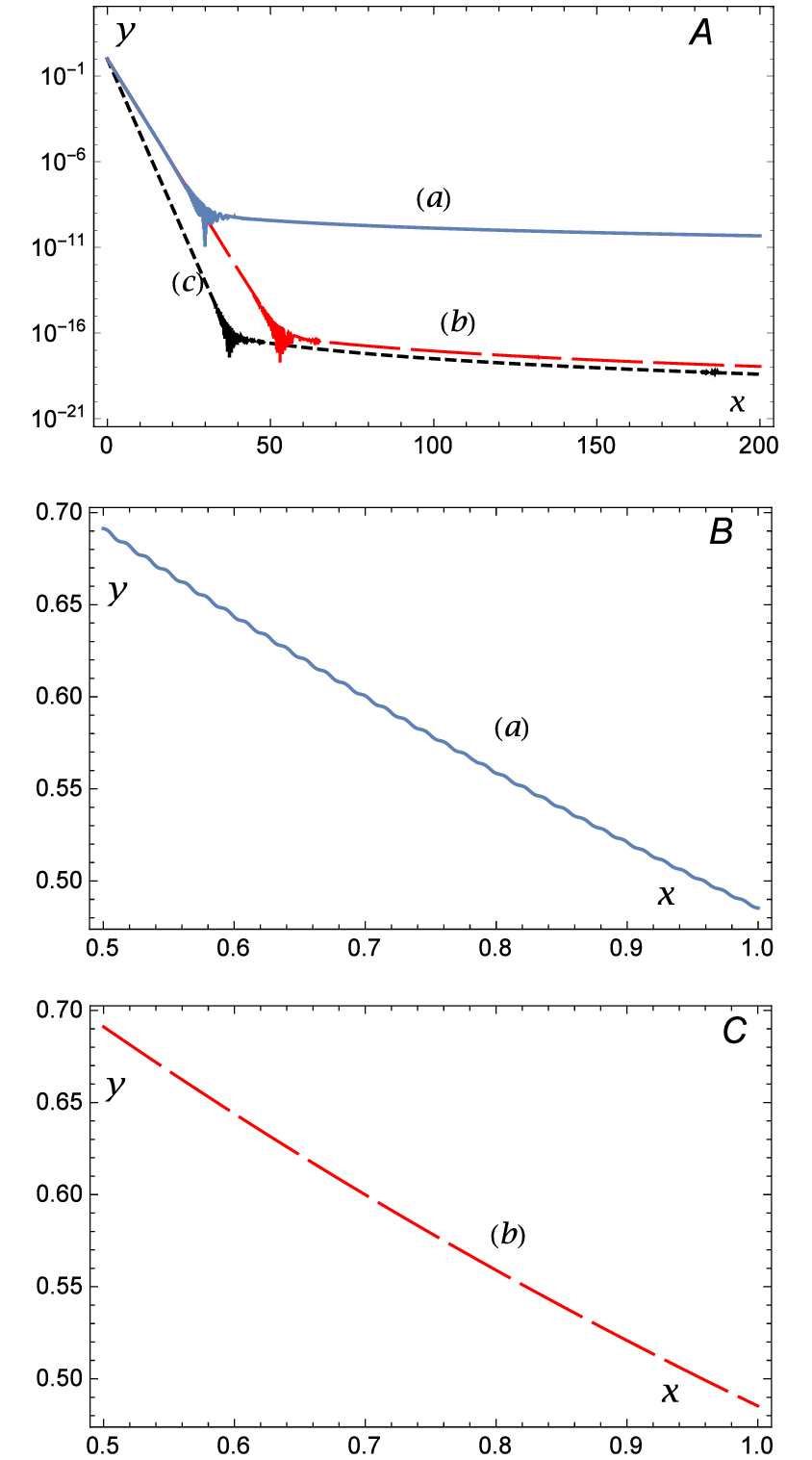}\\
\caption{Decay curves obtained for $\omega(m)$ given by Eq. (\ref{omega-exp}).
Axes: $x =t / \tau_{0} $, and $y$ --- survival probabilities
(panel $A$: the logarithmic scales, $(a)$ the decay curve ${\cal P}_{p}(t)$,
$(b)$ the decay curve ${\cal P}_{0}(t/\gamma)$, $(c)$ the decay
curve ${\cal P}_{0}(t)$; panel $B$:  ${\cal P}_{p}(t)$;
panel $C$: ${\cal P}_{0}(t/\gamma)$ ). The case $\eta {\it\Gamma}_{0}/m_{0} = 0.006$.}
  \label{f4}
\end{center}
\end{figure}

From  an analysis of general properties of the mass (energy) distribution
functions $\omega (m)$  of real unstable systems it follows that
they have properties similar to the scattering amplitude, i.e., they can be decomposed
 into a threshold factor, a pole-function,
with a simple pole (often modeled by $\omega_{BW}(m)$) and a smooth form factor $f(m)$
\cite{fonda,nowakowski3}. This means that $\varrho (m)$ in (\ref{omega-eta}) should have the following form  $\varrho (m) =
\,\omega_{BW} ({m})\,f(m)$, where
$f(m) \rightarrow 0$  as $m\rightarrow \infty$.
Guided by this observation we follow \cite{nowakowski3} and assume that
\begin{equation}
\omega(m) = N\,\sqrt{m - \mu_{0}}\,\frac{{\sqrt{ {\it\Gamma}_{0} }}}{({ m}-{ m}_{0})^{2} +
({{\it\Gamma}_{0}} / {2})^{2}}\,e^{\textstyle{-\eta\,\frac{m}{m_{0} - \mu_{0}}}},
\label{omega-exp}
\end{equation}
with $\eta >0$.  The asymptotic form of the survival amplitude $a_{0}(t)$ for such a $\omega (m)$ is given by
the relation (\ref{a-eta-1/2}). Hence one finds that at late times $t \to \infty$ there is ${\cal P}_{0}(t) \sim 1/t^{3}$  in the case considered.
Decay curves corresponding to $\omega (m)$ defined by  (\ref{omega-exp}) were
find numerically for the case of the particle decaying in the rest system
(the survival probability ${\cal P}_{0}(t)$)  as well as for the moving
particle (the non--decay probability  ${\cal P}_{p}(t)$).
Results are presented in Figs (\ref{f3}) and (\ref{f4}). In order to compare
them with the results obtained for $\omega_{BW}(m)$, calculations were
performed for the same ratios as in that case: $m_{0}/{\it\Gamma}_{0}\, =
\,p/ {\it\Gamma}_{0} = 1000$, and $\mu_{0} = 0$. The ratio $\eta
{\it\Gamma}_{0}/(m_{0} - \mu_{0}) \equiv \eta {\it\Gamma}_{0} /m_{0}$ was
chosen to be $\eta {\it\Gamma}_{0}/m_{0} = 0.01$ (Fig. (\ref{f3})) and
$\eta {\it\Gamma}_{0}/m_{0} = 0.006$ (Fig. (\ref{f4})).

From Figs (\ref{f3}), (\ref{f4}) it is seen that in the case of $\omega (m)
\neq \omega_{BW}(m)$, e.g. when $\omega (m)$ has the form given by Eq.
(\ref{omega-exp}), the survival probabilities ${\cal P}_{0}$ and
${\cal P}_{0}(t/\gamma)$ have an analogous form as the corresponding
probabilities obtained for $\omega (m) = \omega_{BW}(m)$ both for
relatively short times $t \sim \tau_{0}$ and for  long times $t \gg \tau_{0}$.
On the other hand, in the case of the survival probabilities
${\cal P}_{p}(t)$  the difference between decay curves
calculated for the density $\omega(m)$ given by formula (\ref{omega-exp})
and for $\omega (m) = \omega_{BW}(m)$ is significant: The decay curves ${\cal P}_{p}$
calculated for $\omega(m)$ defined in (\ref{omega-exp}) have an oscillating
form at  times $t \sim \tau_{0}$ and shorter  while those obtained for $\omega_{BW}(m)$
do not have.
This is rather unexpected result but it shows
that in the case of moving relativistic particles quantum decay processes
may have nonclassical form even at times shorter than the lifetime.

\section{Analysis of masses and velocities of unstable states}

It was assumed in Sec. 2 and 3 that the momentum $\vec{p}$ of the relativistic unstable particle moving
like a free particle
is conserved. Using this assumption one  usually  concludes that in such a case the velocity
of the particle have to be conserved and constant in time. Such a conclusion is true in the case of
the classical particles: In the case o
a classical object
moving like a free particle
the conservation of the momentum
means that the velocity of this object is constant in time.
 The question is whether such a conclusion is true in the case of moving quantum unstable objects or not.
In order to solve this problem we should analyze relativistic formula for the momentum $\vec{p}$, which within the assumed system of units has the following form: $\vec{p} = m\,\gamma(\vec{v})\,\vec{v}$.
In this relation $m$ is the rest mass of the moving quantum or classical objects and $\vec{v}$ is the velocity of these objects.
From the point of view of the quantum theory the problem is that
the state vector $|\phi_{p}\rangle$ of the form (\ref{phi-p})  corresponding to such a quantum object can not be an eigenvector
of the Hamiltonian $H$ (including the case $\vec{p} = 0$), otherwise it would be that
${\cal P}_{p}(t) =|\langle \phi_{p}|\phi_{p}(t)\rangle|^{2} = |\langle \phi_{p}|\exp\,[-itH]\phi_{p}\rangle|^{2} \equiv 1$ for
all times $t$. The fact that  the vector $|\phi\rangle$ describing the unstable quantum object is not the
eigenvector for $H$ means that the mass (energy) of this object is not defined. Simply
the mass  can not take the exact constant value in this state $|\phi_{p}\rangle$.
In such a case
quantum objects are characterized by the mass (energy) distribution density $\omega (m)$ and
the average mass
\[
<m> = \int_{\mu_{0}}^{\infty}\,m\,\omega(m)\,dm,
\]
 or by the instantaneous mass
(energy) $m_{\phi}(t)$ (see, eg. \cite{ku-plb-2014,ku-dice})  but not by the exact value of the mass.

Let us analyze the properties of the instantaneous mass.
The instantaneous mass $m_{\phi}(t)$ (energy) can be found  using
the exact effective Hamiltonian $h_{\phi}(t)$ governing the time evolution
in the subspace of states spanned by the vector $|\phi\rangle \neq 0$,
\begin{eqnarray}
h_{\phi}(t) &=& \frac{i}{a_{0}(t)}\,\frac{\partial a_{0}(t)}{\partial t}, \label{h(t)-1}\\
&\equiv&  \frac{\langle \phi |H|\phi (t)\rangle}{\langle \phi |\phi (t)\rangle},
\label{h-equiv}
\end{eqnarray}
which results from the Schr\"{o}dinger equation when one looks for the exact
evolution equation for the mentioned subspace of states
(for details see \cite{urbanowski-1-2009,giraldi,ku-plb-2014,ku-dice,ku-2009,pra}).
where the system of units  $\hbar = c =1$ is used.
It is assumed that the vector $|\phi\rangle$ is not an eigenvector of $H$: There does not exist any number $\lambda$ such that
$H|\phi \rangle = \lambda|\phi\rangle$.

Within the assumed system of units the instantaneous mass (energy) of the unstable quantum system in
the rest reference frame is  the real part of $h_{\phi}(t)$:
\begin{equation}
m_{\phi}(t) = \Re\,[h_{\phi}(t)], \label{m(t)}
\end{equation}
and ${\it\Gamma}_{\phi} (t) = - 2\Im\,[h_{\phi}(t)]$ is the instantaneous decay rate.

Using the relation (\ref{h-equiv}) one can find some general properties of $h_{\phi}(t)$ and $m_{\phi}(t)$.
Indeed, if to rewrite the numerator
of the righthand side of (\ref{h-equiv}) as follows,
\begin{equation}
\langle \phi|H|\phi (t)\rangle \equiv \langle
\phi|H|\phi\rangle\,a_{0}(t)\,+\,\langle \phi |H|\phi (t) \rangle_{\perp}, \label{perp}
\end{equation}
where $|\phi (t)\rangle_{\perp} = Q|\phi (t)\rangle$,
$Q = \mathbb{I} - P$ is the projector onto the subspace od decay products,
$P = |\phi\rangle\langle \phi|$ and $\langle \phi|\phi (t)\rangle_{\perp} = 0$,
then one can see
that there is a permanent contribution of
decay products described by $|\phi (t) \rangle_{\perp}$ to the
instantaneous mass (energy) of the unstable state considered.
The intensity of this contribution depends on time $t$.
Using (\ref{h-equiv}) and (\ref{perp}) one finds that
\begin{eqnarray}
h_{\phi}(t) &=& \langle \phi|H|\phi\rangle\, +\, \frac{\langle \phi |H|\phi (t)\rangle_{\perp}}{a_{0}(t)} \label{h-perp-1} \\
&\stackrel{\rm def}{=}& \langle \phi|H|\phi\rangle\,+\,V_{\phi}(t). \label{v-t}
\end{eqnarray}
From this relation one can see that $h_{\phi}(0) = \langle \phi|H|\phi\rangle$ and $V_{\phi}(0) = 0$ if the
matrix elements $\langle \phi|H|\phi\rangle$ exists. It is because
$|\phi (t=0)\rangle_{\perp} =0$ and $a_{0}(t=0)=1$.

\begin{figure}[h!]
\begin{center}
\includegraphics[width=74mm]{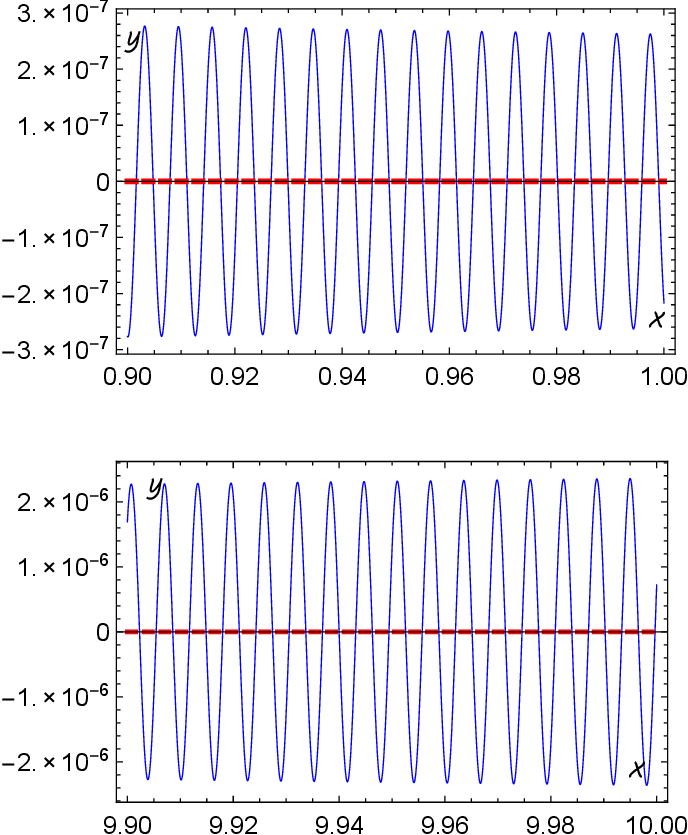}\\
\caption{The instantaneous mass $m_{\phi}(t)$ as
a function of time obtained for $\omega_{BW}(m)$.
Axes: $y = \kappa (t) - 1$, where $\kappa(t)$ is defined  by (\ref{kappa});
$x =t / \tau_{\phi} $: Time is measured in lifetimes.
The horizontal dashed line represents the value of $m_{\phi}(t) = m_{0}$}
  \label{f5}
\end{center}
\end{figure}

\begin{figure}[h!]
\begin{center}
\includegraphics[width=74mm]{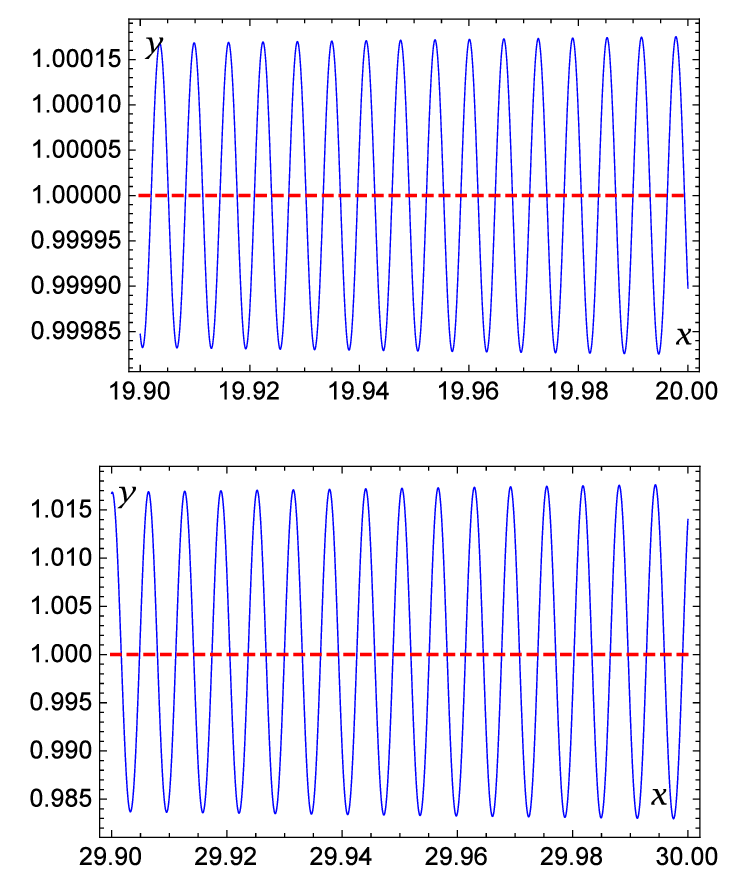}\\
\caption{The instantaneous mass $m_{\phi}(t)$ as
a function of time obtained for $\omega_{BW}(m)$.
Axes: $y = \kappa (t)$, where $\kappa(t)$ is defined  by (\ref{kappa});
$x =t / \tau_{\phi} $: Time is measured in lifetimes.
The horizontal dashed line represents the value of $m_{\phi}(t) = m_{0}$}
  \label{f6}
\end{center}
\end{figure}

\begin{figure}[h!]
\begin{center}
\includegraphics[width=74mm]{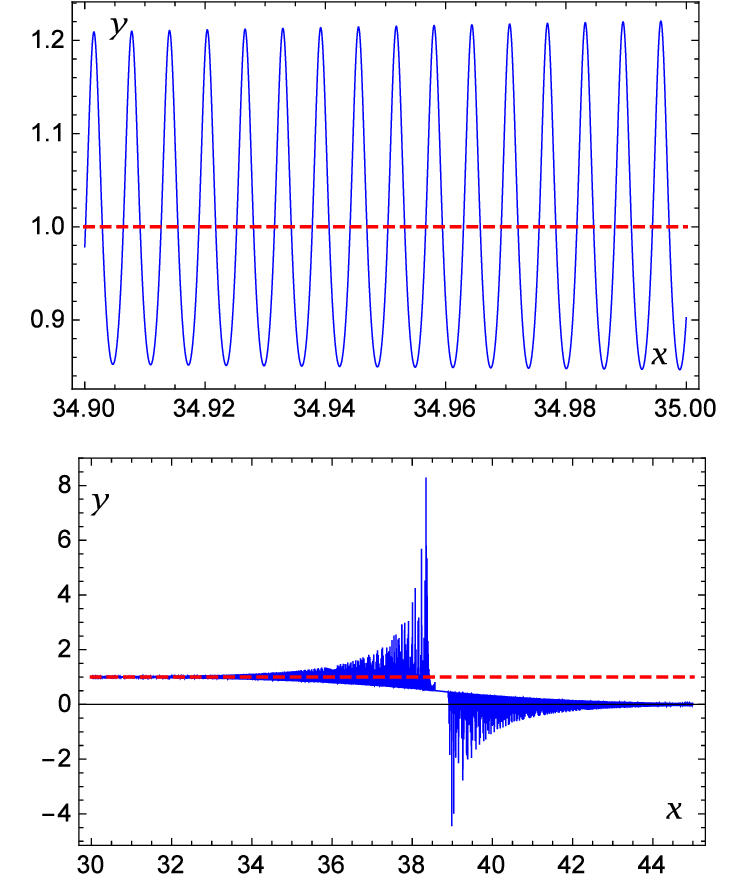}\\
\caption{The same as in Fig (\ref{f6}) for longer times.}
  \label{f7}
\end{center}
\end{figure}

Now let us  assume  that $\langle \phi|H|\phi\rangle$ exists and
$i \frac{\partial a_{0}(t)}{\partial t} \equiv \langle \phi| H|\phi;t\rangle$ is a continuous function of time $t$ for $0 \leq t < \infty$.
If these assumptions are satisfied then $h_{\phi}(t)$ is a continuous function of time $t$ for $0 \leq t < \infty$ and $h_{\phi}(0) =
\langle \phi|H|\phi\rangle$ exists.
Now if to  assume
that for
$0\leq t_{1} \neq t_{2}$ there is $h_{\phi}(0) = h_{\phi}(t_{1}) = h_{\phi}(t_{2}) = const$
then from the continuity of $h_{\phi}(t)$ immediately follows that
there should be $h_{\phi}(t) = h_{\phi}(0) \equiv \langle \phi|H|\phi\rangle = const$ for any $t \geq 0$.
Unfortunately such an observation contradicts implications of  (\ref{h-perp-1}), (\ref{v-t}):
From the relations
(\ref{h-perp-1}), (\ref{v-t}) one concludes that it is possible if, and only if, 
\begin{equation}
V_{\phi}(t > 0) = 0, \label{v-t>0}
\end{equation}
for every $t$ such that $0<t<\infty$. There is $|a_{0}(t)| > 0 $ for $t < \infty$, therefore
\begin{equation}
V_{\phi}(t > 0) = 0 \;\;\Leftrightarrow\;\; \langle \phi |H|\phi (t>0)\rangle_{\perp} = 0, \label{f-perp-f}
\end{equation}
for every $t  >0$ and $ t < \infty$. The relation (\ref{f-perp-f}) can take place if, and only if,
\begin{equation}
( \langle \phi| H)^{+} \equiv H |\phi\rangle \;\;\perp\;\; |\phi (t>0)\rangle_{\perp}\;\;\;\;{\rm for\;all}\;\;t > 0.
\end{equation}
This last condition leads to the conclusion that
\begin{equation}
\big\{|V_{\phi}(t>0)| = 0\;\; {\rm for\;every}\;\;\;t >0 \big\}\;\;  \;\Leftrightarrow\;\;\; H|\phi \rangle = \lambda |\phi \rangle.
\end{equation}
This observation means that
\begin{equation}
h_{\phi}(t) = const,
\end{equation}
if and only if there is no any decay of the state $|\phi \rangle$ considered (if there is no any transitions between ${\cal H}_{\parallel} = P {\cal H}$ and ${\cal H}_{\perp}$). So, in the case of unstable systems
$h_{\phi}(t>0) \neq const$, which means that in the case of unstable systems the instantaneous mass (energy) $m_{\phi}(t) \equiv \Re\,[h_{\phi}(t)]$ and the instantaneous decay rate ${\it\Gamma}_{\phi}(t)$ can not be constant in time: $m_{\phi}(t) \neq const$ and ${\it\Gamma}_{\phi}(t) \neq const$.
Results of numerical calculations presented in Figs (\ref{f5}) --- (\ref{f7}) (or those one can find in \cite{ku-plb-2014,ku-dice}) confirm this conclusion.
In Figs (\ref{f5}) --- (\ref{f7})
the function,
\begin{equation}
\kappa (t) = \frac{m_{\phi}(t) - \mu_{0}}{m_{0} - \mu_{0}}, \label{kappa}
\end{equation}
is presented, which illustrates a typical form of time--varying  $m_{\phi} (t)$.
(All calculations were performed for $(m_{0} - \mu_{0})/{\it\Gamma}_{0} = 1000$).

As it is seen from  Figs (\ref{f5}), (\ref{f6}), (\ref{f7})  the amplitude
 of variations of $m_{\phi}(t)$ needs not be large at relatively short times:
It is almost negligible small
but these variations always exist (see Figs (\ref{f5}) --- (\ref{f7}) and results presented in \cite{ku-2015}).
When the time increases the amplitude of these variations grows  and reaches maximal values for times $t \sim T$.
Now if this particle is a moving relativistic particle
then within the assumed system of units its momentum equals
$\vec{p} = m_{\phi}\,\gamma(\vec{v})\,\vec{v}$, where $m_{\phi}$ is the rest mass of the particle $\phi$, $\vec{v}$ is the velocity.
The total  momentum (and energy)  of the
objects moving like a free particle
both quantum and classical must be conserved.
Thus it has to be $\vec{p}(t_{1})=\vec{p}(t_{2})$, that is $m_{\phi}(t_{1})\,\gamma(\vec{v})\,\vec{v} =
m_{\phi}(t_{2})\,\gamma(\vec{v})\,\vec{v}$ for any $t_{1} \neq t_{2}$. It is possible only if  changes of
$m_{\phi}(t)$ are compensated by suitable changes of $\gamma (\vec{v})\,\vec{v}$,
that is by corresponding  changes in the velocity $\vec{v}$.
(A similar  mechanism  was described in \cite{ku-plb-2014,ku-dice}, where its consequences were analyzed
for times of the order of the cross--over time $T$).
So the principle of conservation of the momentum  forces
compensation of changes in the instantaneous mass $m_{\phi}(t)$
through appropriate changes in the velocity of the moving unstable system.
(It is a pirouette like effect).
This is why the assumption $\vec{v} = const$ when considering moving quantum unstable objects
leads to the result $P_{\vec{v}}(t) = P_{0}(\gamma t)$, i.e., to the result never observed in experiments \cite{shirkov1}.
Thus the assumption  $\vec{p} = const$ mentioned seems to be the only acceptable choice in the case of moving
quantum unstable systems (see also a discussion in \cite{giacosa1}).

Let us analyze now implications of the observation that the velocity $\vec{v}$ of the
quantum unstable system moving like a free particle
 can not be constant in time
and it has to vary in time $\vec{v} \equiv \vec{v}(t) \neq const$. This property has an effect that $\frac{d \vec{v}}{dt} \neq 0$. Now let us denote
by ${\cal O}'$
the reference frame
which moves together with the moving quantum unstable system considered and
in which this system is in rest.
This reference frame
moves relative to ${\cal O}$ with the velocity $\vec{v} = \vec{v}(t) \neq const$ measured in ${\cal O}$.
The observation that $\frac{d \vec{v}}{dt} \neq 0$ means
that  the rest reference frame ${\cal O}'$ of the
quantum unstable
system moving like a free particle
can not be the inertial one.

\section{Discussion and Conclusions}

Let us begin from a general remark:
In any case we should remember that
the relation (\ref{a-p=a-0}) is the classical physics relation
and that the quantum decay processes are analyzed in this paper.
The relativistic time--dilation relation
in its  form known from classical physics does not
need to  manifest itself
in quantum processes in the same way as in classical physics processes.
It is also important to be aware that as it was shown in \cite{giacosa2} the Quantum Field Theory
models of the decay processes can be also described within  the formalism used in Sec. 2.

All results presented in Figs (\ref{f1}) --- (\ref{f4}) show decay curves seen
by the observer ${\cal O}$ in his rest reference frame (curves ${\cal P}_{0}(t/\gamma)$
correspond to the situation when the classical dilation relation (\ref{a-p=a-0}) is
assumed to be true in the case of quantum decay processes).
The time (the horizontal axes) in all these
figures is the time measured by the observer in his rest system.
These results show that
Stefanovich--Shirokov theory \cite{stefanovich,shirkov}
predict a such form of the survival probability ${\cal P}_{p}(t)$
that the expected relation (\ref{a-p=a-0}) holds to very good approximation
only for times $t \sim \tau_{0}$ and only for $\omega (m) = \omega_{BW}(m)$.
The  visible difference  between ${\cal P}_{p}(t)$ and ${\cal P}_{0}(t/\gamma)$
takes place at times $ t \gg \tau_{0}$ but this needs not mean that this theory is wrong:
To this day there have been no published reports on experiments analyzing the form of the decay law
of moving relativistic unstable  particles at times $t \gg \tau_{0}$ or $t \sim T$ and $t > T$.

Analyzing the results presented in Figs
(\ref{f1}) --- (\ref{f4}) we can conclude that properties of
the survival probability of the moving unstable
particle, ${\cal P}_{p}(t) = |a_{p}(t)|^{2}$,
where $a_{p}(t)$ is calculated using the Eq.
(\ref{a-p}) (i.e the formula derived in \cite{stefanovich,shirkov}), are much more sensitive to the form of
$\omega (m)$ than properties of ${\cal P}_{0}(t)$.
It is a general observation. Another general conclusion
following from these results is that starting from times $t$
from the transition time region, $t > T$, the decay process
of moving particles is much slower than
one would expect assuming the standard dilation relation (\ref{a-p=a-0}).

From Figs (\ref{f3}) and (\ref{f4}) it follows that in the case of moving
relativistic unstable particles the standard relation (\ref{a-p=a-0})
does not apply in the case of the density $\omega (m)$ of the form (\ref{omega-exp})
and leads to the wrong conclusions for such densities.
Results presented in these Figures  show also that a conclusion drawn
in \cite{stefanovich,shirkov,ku-2014} on the basis of studies of the model defined by
the Breit--Wigner density $\omega_{BW}(m)$
that the relation (\ref{a-p=a-0}) is
valid for not more than few lifetimes is true only for the density $\omega_{BW}(m)$
and need not be true for densities $\omega (m)$ having a more general form.
Similar limitations concern the result presented in \cite{exner}, where it is stressed that
the approximations used to derive the final result may work only for times no longer than a few lifetimes.
What is more, a detailed analysis shows that the final result presented therein was obtained
using the non--relativistic limit of $\sqrt{m^{2} + p^{2}}$. There was used the  following approximation:
$\sqrt{m^{2} + p^{2}} \simeq m + \frac{p^{2}}{2m} +\ldots$
(see \cite{exner}, formula (20) and then (30a), (30b)).
So, in general the relation (\ref{a-p=a-0}) can be considered as sufficiently
accurate approximation only for not too long times $t$ if at these times
${\cal P}_{p}(t)$ has the same  exponential form as the decay laws obtained
within classical physics considerations. If quantum effects force
${\cal P}_{p}(t)$ to behave non--classically at these times then the
relation (\ref{a-p=a-0}) which is the classical physics relation  is not applicable.

In general,  as it follows from the results obtained within the considered theory and presented in  Figs. (\ref{f3}),
(\ref{f4})  contrary the standard expectations based on the  classical physics time dilation
relation of the special relativity
some quantum effects  should be registered
earlier  by the observer ${\cal O }$ studying the behavior of moving
unstable particles in relation to his rest reference frame
than the same effects observed by ${\cal O}$ in the case of the particles
decaying in the common rest reference frame for the particle and  the observer
${\cal O}$: The transition times region, that is the time region when contributions
from the exponential and late time non--exponential parts of the amplitude
$a_{p}(t)$ or $a_{0}(t)$ are comparable, which manifest itself as a sharp and
frequent oscillations of the survival probability, takes place earlier for
${\cal P}_{p}(t)$ (Fig. (\ref{f2}), the curve $(a)$ and Figs. (\ref{f3}) and
(\ref{f4}), panel $A$, curves $(a)$)
than for ${\cal P}_{0}(t)$ (Fig. (\ref{f2}), the curve $(c)$
and Figs. (\ref{f3}) and (\ref{f4}), panel $A$, curves $(c)$).
The same observation concerns results
presented in panels $B$ of Figs. (\ref{f3}) and (\ref{f4}).

These properties that is
the form of the decay curves presented in panel A of Figs  (\ref{f1}), (\ref{f3}) and (\ref{f4}) can be easy explained
analyzing the equivalent expression of the formula (\ref{a-p}) for $a_{p}(t)$:
\begin{eqnarray}
a_{p}(t)
&\equiv& \int_{\mu_{0}}^{\infty} \omega(m)\;
e^{\textstyle{-\,i\,m \,\gamma_{m}\,t}}\,d{m} = a_{<p}(t) + a_{>p}(t), \label{a-p-g}
\end{eqnarray}
where $\gamma_{m}$ can be equivalently written as $\gamma_{m} \equiv \sqrt{1 + \frac{p^{2}}{m^{2}}}$ and within the used system of units
\begin{eqnarray}
a_{<p}(t)
&\equiv& \int_{\mu_{0}}^{p} \omega(m)\;
e^{\textstyle{-\,i\,m \,\gamma_{m}\,t}}\,d{m},  \label{a-p-<}\\
a_{>p}(t)
&\equiv& \int_{p}^{\infty} \omega(m)\;
e^{\textstyle{-\,i\,m \,\gamma_{m}\,t}}\,d{m},  \label{a-p->}
\end{eqnarray}
It is easy to see that for  $m < p$ there is $\gamma_{m} > \sqrt{2}$ and   $\gamma_{m}$ becomes very large for $m \ll p$, which means that
$a_{<p}(t)$ reaches values proper for times $t$  of order of the crossover time $T$  much earlier comparing with $a_{0}(t)$ given by formula (\ref{a-spec}).
Therefore the visible oscillations of decay curves of moving particles can begin earlier than in case of the particles decaying in the rest system.
On the other hand for $m > p$ one has   $\gamma_{m} < \sqrt{2}$ and for $m \gg p$ we observe that  and $\gamma_{m} \simeq 1$ which shows that contribution of $a_{>p}(t)$ into $a_{p}(t)$ is almost the same as in the case the of $a_{0}(t)$, which explains why at very late time the decay curves of moving unstable particles  presented in panels A of (\ref{f1}), (\ref{f3}) and (\ref{f4}) have the same form as in case of particles decaying in the rest system. The final form of the decay curve ${\cal P}_{p}(t)$ of the moving unstable particle with a definite momentum depends on the balance of contributions to $a_{p}(t)$ coming from
amplitudes $a_{<p}(t)$ and $a_{>p}(t)$ and on the interference between them:
\begin{equation}
{\cal P}_{p}(t) = |a_{p}(t)|^{2} \equiv |a_{<p}(t) \,+\,a_{>p}(t)|^{2}. \label{P-p><}
\end{equation}
The balance between contributions of $|a_{<p}(t)|^{2}$ and $|a_{>p}(t)|^{2}$ into ${\cal P}_{p}(t)$ depends on the form and properties of $\omega (m)$.

In all figures the time is measured in lifetimes $\tau_{0}$.
So, fluctuations of ${\cal P}_{p}(t)$
calculated for the density $\omega (m) = \omega_{BW}(m)$ and
presented in Fig (\ref{f2}) (the decay curve $(a)$)
are rather unmeasurable: They take a place at $t \sim 20 \tau_{0}$.
On the other hand
similar fluctuations appearing in the case of $\omega(m)$
given by Eq. (\ref{omega-exp}) and presented
in Fig. (\ref{f3}) (panel $B$, the decay curve $(a)$)
and Fig. (\ref{f4}) (panel $B$) take place
at times $ t\leq \tau_{0}$ and longer. This means that
the probability that they can be registered
in some cases is very high.

Results of Sec. 4 explain the growing with time differences between ${\cal P}_{p}(t)$ and ${\cal P}_{0}(t/\gamma)$.  Note that  ${\cal P}_{0}(t/\gamma)$ corresponds to the  classical physics expectations.  The cause of these differences is pure quantum effect: Fluctuations in time of the instantaneous mass $m_{\phi}(t)$ of the unstable quantum system. Simply for relatively long times fluctuations of this instantaneous mass $m_{\phi}(t)$
become significant and grows with time $t$. Hence variations of $\vec{v}(t)$ have to be larger and larger. This means that deviations from the classical physics predictions become also large and grows with the increasing fluctuations of $m_{\phi}(t)$.

Let us take a look again at Figs. (\ref{f3}) and (\ref{f4}).
A more detailed analysis of panels $B$ in the Figs. (\ref{f3}) and (\ref{f4}) indicates
striking similarity of the decay curves ${\cal P}_{p}(t)$ presented there by
solid lines (curves ${\cal P}_{0}(t/\gamma)$ are represented there by long--dashed lines)
to the results  presented in Figs. 3 -- 5 in \cite{litvinov1} known as
the "GSI anomaly".
This suggests that the nature of GSI anomaly is probably purely quantum-mechanical.
(Readers can meet a several
theoretical proposals that attempt to explain the GSI anomaly:
There are authors using the interference of two mass eigenstates
(see, eg. \cite{kienert}); Some authors use neutrino oscillations \cite{ivanov};
In \cite{madrid} time is used as a dynamical variable and the time
representation is used; In \cite{giacosa} a truncated Breit--Wigner mass
distribution with an energy--dependent decay with ${\it\Gamma}$
such that $\omega (m) = 0$ for $m< \Lambda_{1}$ and $\omega (m) = 0$
for $m > \Lambda_{2} > \Lambda_{1}$  is applied,
and so on).

One more observation.
Note that from properties of the relativistic
expression ${\cal P}_{p}(t)$ it follows that
within the considered theory
the number
of unstable particles which are able to survive
up to times $t$ longer than the transition time $T$ is much greater
then one would expect performing suitable estimations
using ${\cal P}_{0}(t/\gamma)$ (see results presented in
panels A of Figs (\ref{f1}), (\ref{f3}) and (\ref{f4})) and that
the decay process at times $t > T$ is significantly slower
than it results from the properties of  ${\cal P}_{0}(t/\gamma)$. These
properties seem to be important when one analyzes some accelerator
experiments with unstable particles of high energies
or results of observations of
some astrophysical and cosmological process:
In many astrophysical
processes an extremely huge numbers of unstable particles are created and
they all are moving with relativistic velocities.
These numbers are so huge that many of them may survive up to  times $t \sim T$ or
even to much longer times $t \gg T$.
So, taking into account results presented in panels A of the above mentioned
Figures one can conclude that at asymptotically late times $t > T$ much
more unstable particles
may be found undecayed than
an observer from Earth
expects considering the classical relation (\ref{a-p=a-0}).

All the above conclusions following from the results presented in Figs (\ref{f1}) --- (\ref{f3})
are the consequence of the form of the amplitude $a_{p}(t)$ derived in \cite{stefanovich, shirkov} and briefly described in
Sec. 2. The question is if this amplitude reflects correctly real properties of the moving unstable quantum objects
(particles) and thus if the possible  effects predicted using this $a_{p}(t)$ and described in this Section
can occur: Only the suitable experiments can decide about this.
The problem is that
all known tests of the relation (\ref{a-p=a-0}) were performed
for times $t \sim \tau_{0}$ (where $\tau_{0}$ is the lifetime)
(see, eg. \cite{muon1,muon2}).
In the light of the results presented in this paper and of the
above discussion  one concludes that the problem of the fundamental importance is to
examine how the relativistic dilation really works in quantum decay
processes of moving relativistic particles for very long times: From
times longer than a few lifetimes up to times longer than the cross--over
time $T$. Only this kind of an experiment can decide how time dilation
being classical physics relation is manifested in the quantum decay processes of relativistic particles.

\end{document}